\documentclass[smallextended]{svjour3}

\usepackage{epsfig}
\usepackage{amsfonts}
\usepackage{amsmath}
\usepackage{amssymb}
\usepackage{bm}
\usepackage{graphicx}

\newcommand{\simle}{\hspace*{0.2em}\raisebox{0.5ex}{$<$}
     \hspace{-0.8em}\raisebox{-0.3em}{$\sim$}\hspace*{0.2em}}

\newcommand{\g}{\gamma}

\newcommand{\la}{\lambda}
\newcommand{\simu}{\sigma^{\mu\nu}}
\newcommand{\Fmu}{F_{\mu\nu}}
\newcommand{\Gmu}{G^a_{\mu\nu}}

\newcommand{\slashT}{\slash\hspace{-0.5em}T}
\newcommand{\slashTsub}{\slash\hspace{-0.4em}T}

\newcommand{\slashPT}{\slash\hspace{-0.6em}P\slash\hspace{-0.5em}T}
\newcommand{\slashPTsub}{\slash\hspace{-0.45em}P\slash\hspace{-0.4em}T}

\newcommand{\qb}{\bar q}
\newcommand{\Nb}{\bar N}
\newcommand{\Fp}{F_\pi}

\newcommand{\tb}{\bar \theta}

\newcommand{\mpi}{m_{\pi}}
\newcommand{\MQCD}{M_{\mathrm{QCD}}}

\newcommand{\Or}{\mathcal O}

\newcommand{\dslash}[1]{#1 \llap{/\kern-0.5pt}}
\newcommand{\Dslash}[1]{#1 \llap{/\kern+1.2pt}}
\newcommand{\DDslash}[1]{#1 \llap{/\kern+2.3pt}}
\newcommand{\dslashh}[1]{#1 \llap{/\kern+1pt}}

\newcommand{\boldtau}{\mbox{\boldmath $\tau$}}
\newcommand{\boldpi}{\mbox{\boldmath $\pi$}}

\newcommand{\CP}{C\hspace{-.5mm}P}

\begin{document}

\title{Parity- and Time-Reversal-Violating Moments of Light Nuclei}
\author{Jordy de Vries}
\institute{KVI, theory group\at
              Zernikelaan 25, 9747 AA Groningen, The Netherlands \\
              \email{devries@kvi.nl} }   
\date{July 30, 2012}

\maketitle

\begin{abstract}
I present the calculation of parity- and time-reversal-violating moments of the nucleon and light nuclei, originating from the QCD $\bar \theta$ term and effective dimension-six operators. By applying chiral effective field theory these calculations are performed in a unified framework. I argue that measurements of a few light-nuclear electric dipole moments would shed light on the mechanism of parity and time-reversal violation.

\end{abstract}

\section{Introduction}
Apart from the direct search for new physics in high-energy collider experiments, complementary searches are taking place by performing precision measurements at a much lower energy scale. The focus of this talk is one of these precision experiments, the ongoing and planned searches for parity- and time-reversal-violating ($\slashPT$) moments. 
The search for electric dipole moments (EDMs) is interesting because the accuracy is high enough to probe energy scales where physics beyond the Standard Model (SM) is expected to appear, but not high enough to actually reach the EDM predictions from the known source of $\CP$ violation in the quark-mixing matrix. That is, with current experimental accuracy EDMs are ``background-free" probes of unmeasured sources of $P$ and $T$ violation. Hadronic and nuclear EDMs are not SM-free probes because a finite measurement could still be due to $\slashPT$ in the strong interaction parametrized by the QCD vacuum angle $\tb$ \cite{'tHooft:1976up}. The current neutron EDM limit \cite{dnbound}, however, limits $\tb$ to be very small, $\tb < 10^{-10}$ \cite{Cre79}, such that there is room for $\slashPT$ sources from physics beyond the SM. In fact, the existence of such a source is well motivated by the universal asymmetry between matter and antimatter.

In the last decade plans have been made to measure the EDMs of the proton, deuteron, and helion (nucleus of ${}^3\mathrm{He}$) directly in storage rings with an expected accuracy exceeding that of the current neutron EDM limit by two to three orders of magnitude \cite{storageringexpts}. One can think of measurements of the EDMs of other light nuclei such as the triton (nucleus of ${}^3\mathrm{H}$) as well. This talk focuses on these proposed measurements. In particular the question whether we can extract from them the fundamental $\slashPT$ source. Some other interesting observables are discussed as well. 

\section{Effective field theories}

If we assume that new $\slashPT$ originates in a scale considerably higher than the electroweak scale, we can treat the SM as an effective field theory (EFT) and add higher-dimensional operators consisting of SM fields and obeying the SM gauge symmetries. The most important $\slashT$ operators start at dimension six \cite{dim6origin} and are suppressed by two powers of a high-energy scale, $M_{\slashTsub}$, where the new $\slashPT$ physics originates. At the electroweak scale around $M_W$, the mass of the W boson, these operators can contain all SM fields. We lower the energy to a scale $\MQCD \simeq 1\,\,\mathrm{GeV}$ and keep only effective $\slashPT$ operators in terms of light quarks, gluons, and photons (we will neglect the leptons). The effective dimension-four and -six $\slashPT$ Lagrangian around $\MQCD$ is given by 
\begin{eqnarray}\label{QCDscale}
\mathcal L_{4\mathrm{-}6} &=&\tb \frac{\bar m}{2} (1-\varepsilon^2)\, \qb i \g^5 q-\frac{i}{2}\qb \left(d_0+d_3 \tau_3\right)\simu \g_5 q \; \Fmu \nonumber\\
&&-\frac{i}{2}\qb \left(\tilde{d}_0+\tilde{d}_3 \tau_3\right)
                   \simu\g_5\la^a q \; \Gmu + \frac{d_{W}}{6} f^{a b c} \varepsilon^{\mu \nu \alpha \beta} G^a_{\alpha \beta} G_{\mu \rho}^{b} G^{c\, \rho}_{\nu} 
\nonumber\\
&&+ \frac{1}{4} \textrm{Im}{\Sigma_1} \left( \bar q q\, \bar q i \gamma^5 q - \bar q \boldtau q\, \cdot \bar q \boldtau i \gamma^5 q \right)
\nonumber\\
&&+  \frac{1}{4} \textrm{Im}{\Sigma_8} \left( \bar q \lambda^a q\, \bar q i \gamma^5 \lambda^a q - \bar q \boldtau \lambda^a q\, \cdot \bar q \boldtau i \gamma^5 \lambda^a q \right),
\end{eqnarray}
in terms of the light-quark doublet $q = (u\,\,d)^T$, the photon and gluon field strengths $F_{\mu\nu}$ and $G_{\mu\nu}^a$, $\boldtau$ the Pauli matrices in isospin space, the Gell-Mann matrices $\lambda^{a}$ in color space and the associated
structure constants $f^{abc}$, the average light-quark mass $\bar m = (m_u + m_d)/2$ and quark-mass difference $\varepsilon =(m_d-m_u)/2\bar m$. The first operator in Eq. \eqref{QCDscale} is the SM $\tb$ term after using an axial $U(1)$ rotation to move all $\slashPT$ into the quark mass and after vacuum alignment \cite{Fujikawa:1979ay}. The remaining terms in Eq. \eqref{QCDscale} are dimension-six operators. The second and third denote, respectively, the light-quark EDMs and chromo-EDMs (CEDMs). Although they have canonical dimension five, in order to conserve $SU_L(2)$ gauge symmetry they are coupled to the Higgs field at the electroweak scale \cite{Rujula}. Because the EDMs and CEDMs flip the chirality of the quarks we assume the interactions to be proportional to the light-quark Yukawa couplings. 
We therefore write
\begin{equation}
d\sim\mathcal{O}\left(e\delta
\frac{\bar{m}}{M_{\slashT}^{2}}\right)\ ,
\qquad\tilde{d}\sim \mathcal{O}\left(4\pi\tilde{\delta}
\frac{\bar{m}}{M_{\slashT}^{2}}\right)\ ,
\end{equation}
in terms of the proton charge $e$ and the dimensionless parameters $\delta$ and $\tilde \delta$. The fourth
term in Eq. \eqref{QCDscale} is the gluon CEDM (gCEDM) \cite{Weinberg:1989dx}, the fifth and sixth terms $\slashPT$ four-quark (FQ) operators  \cite{dim6origin}. These terms have coefficients
\begin{equation}
d_{W}\sim\mathcal{O}\left(\frac{4\pi w}{M_{\slashT}^{2}}\right),\qquad \textrm{Im}\Sigma_{1,8}=
\mathcal{O}\left(\frac{(4\pi)^{2}\sigma_{1,8}}{M_{\slashT}^{2}}\right)\ ,
\label{w}
\end{equation}
in terms of dimensionless parameters $w$ and $\sigma_{1,8}$. The
sizes of $\delta$, $\tilde\delta$, $w$, and $\sigma_{1,8}$
depend on the mechanisms of electroweak and $PT$ breaking and
on the running to low energies where nonperturbative QCD sets in.

A measurement of any hadronic or nuclear EDM can be fitted by all of the sources in Eq. \eqref{QCDscale}. An important open question and the focus of this talk is whether it is possible to identify the fundamental $\slashPT$ source from several measurements of nucleon and light-nuclear EDMs. Can we separate the QCD $\tb$ term from beyond-the-SM physics? If so, can we also differentiate between the various dimension-six operators?  

To answer these questions we need to calculate $\slashPT$ observables in terms of the operators in Eq. \eqref{QCDscale}. This is a difficult problem since it involves nonperturbative QCD. Here we use chiral perturbation theory ($\chi$PT), the low-energy (below $\MQCD$) EFT of QCD. In $\chi$PT, instead of quarks and gluons, the effective degrees of freedom are pions and nucleons whose interactions are dictated by the symmetries of QCD and how they are (spontaneously) broken. A particular important role is played by the approximate chiral symmetry of
QCD, $SU_{L}(2)\times SU_{R}(2)\sim SO(4)$. Since it is not manifest in the hadronic
spectrum (there exists no partner of the nucleon with odd parity), which does have an approximate isospin symmetry, 
chiral symmetry must be spontaneously broken down to the isospin subgroup 
$SU_{L+R}(2) \sim SO(3)$. The corresponding Goldstone bosons 
are identified with the pions whose mass depend on the chiral-symmetry-breaking quark mass, $\mpi^2 =\Or(\bar m \MQCD)$. Although an expansion in the strong coupling constant is lost, $\chi$PT brings in a new expansion parameter $Q/\MQCD$ where $Q\sim \mpi$ is the typical energy of the process. Each hadronic interaction is associated with a low-energy constant (LEC) whose size can be estimated by various techniques such as QCD sum rules \cite{Pospelov:2005pr} or naive dimensional analysis (NDA) \cite{NDA}. Here we use NDA. Ideally, these estimations are replaced by lattice-QCD calculations.

While they all break $P$ and $T$, the dimension-four and -six
operators break chiral symmetry differently from each other and they give rise to different effective interactions. Therefore, each $\slashPT$ source generates  a characteristic pattern of relations between different observables. The observation of such a pattern in, for example, nucleon and nuclear EDM experiments would direct us to the dominant $\slashPT$ source at the QCD scale. Once this is known, the next step would be to infer the dominant source at the electroweak scale.

The best way to illustrate this idea is by looking at a subset of the effective $\slashPT$ chiral Lagrangian consisting of two $\slashPT$ pion-nucleon ($\pi N$)  interactions
\begin{eqnarray}\label{g0g1}
\mathcal L_{\slashT, \pi N}&=& -\frac{\bar g_0}{\Fp} \Nb \boldtau\cdot\boldpi N -\frac{\bar g_1}{\Fp} \pi_3 \Nb\! N, 
\end{eqnarray}
in terms of the nucleon doublet $N = (p\,\,n)^T$, the pion triplet $\boldpi$, and the pion decay constant $\Fp = 186$ MeV. Both interactions in Eq. \eqref{g0g1} break chiral symmetry, but only $\bar g_1$ breaks isospin symmetry. The $\tb$ term, a chiral-breaking and isospin-conserving interaction, at low energy induces $\bar g_0$ directly, but in order to generate $\bar g_1$ isospin breaking needs to be brought in by the quark-mass difference. Applying NDA we find the following scalings \cite{BiraEmanuele}
\begin{equation}
\bar{g}_{0}=\Or\!\left(\bar{\theta}\frac{m_{\pi}^{2}}{\MQCD}\right),
\qquad
\bar{g}_{1}=\Or\!\left(\varepsilon\bar{\theta}\frac{m_{\pi}^{4}}{\MQCD^{3}}
\right),
\label{NDAtheta}
\end{equation}
such that $\bar g_1$ enters in the next-to-next-to-leading order (NNLO) Lagrangian. 

On the other hand the dimension-six qCEDM has an isovector component such that $\bar g_0$ and $\bar g_1$ are expected to be of similar size.
In principle, the same holds for the qEDM but for this source a photon needs to be integrated out in order to generate the $\pi N$ couplings. The additional cost of $\alpha_{\mathrm{em}}/4\pi$ causes the resulting interactions to be subleading with respect to other $\slashPT$ interactions containing explicit photons. 

Finally, the gCEDM and the two FQ operators conserve chiral and isospin symmetry. For simplicity we refer to them as chiral-invariant ($\chi$I) sources
and use $w$ to denote both $w$ and $\sigma_{1,8}$: 
$
\{w,\sigma_{1},\sigma_{8}\}\to w.
$
In order for these sources to generate the chiral-breaking $\pi N$ interactions, the quark mass (difference) needs to be inserted which suppresses the interactions.
As a consequence the nonderivative $\pi N$ interactions in Eq. \eqref{g0g1} appear at the same order as $\chi$I two-derivative $\pi N$ interactions. More importantly, $\chi$I nucleon-nucleon ($N\!N$) interactions appear at leading order (LO) in the $\slashPT$ $N\!N$ potential \cite{TVpotential}. 

Apart from $\slashPT$ $\pi N$ interactions there is no reason to not include other kinds of interactions. The full chiral Lagrangian for the $\tb$ term has been constructed in Ref. \cite{BiraEmanuele} and for the dimension-six sources in Ref. \cite{dim6}. It is found that at LO the EDMs of the nucleon and light nuclei depend on six $\slashPT$ interactions \cite{Vri12}
\begin{eqnarray}
\mathcal{L}_{\slashPTsub} & = & 
-\frac{1}{F_{\pi}}
\bar{N}\left(\bar{g}_{0}\,\boldtau\cdot\boldpi+\bar{g}_{1}\pi_{3}\right)N - 2\, \Nb\left(\bar{d}_{0}+\bar{d}_{1}\tau_{3}\right)S^{\mu}N\, v^{\nu}\Fmu
\nonumber \\
& & +\bar{C}_{1}\bar{N}N \, \partial_{\mu}(\bar{N}S^{\mu}N)
+\bar{C}_{2}\bar{N}\boldtau N\cdot
\partial_{\mu}(\bar{N}S^{\mu}\boldtau N).
\label{6Lecs}
\end{eqnarray}
Here we work in the 
heavy-baryon framework \cite{Jenkins:1990jv}
where, instead of gamma matrices, it is the nucleon 
velocity $v^\mu$ and spin $S_\mu$
that appear. The first two interactions in Eq. \eqref{6Lecs} are the $\slashPT$ $\pi N$ interactions discussed above. The third and fourth terms are short-range contributions to, respectively, the isoscalar and isovector nucleon EDM. The last two terms are $\chi$I $\slashPT$ $N\!N$ interactions. 

Which of these six interactions is relevant depends on the observable
we are calculating and on the fundamental $\slashPT$ source. For $\tb$ and qCEDM the $\slashPT$ $\pi N$ and nucleon-photon ($N \g$) interactions need to be taken into account, although for $\tb$, as explained above, $\bar g_1$ is small compared to $\bar g_0$. For these sources $N\!N$ interactions enter at higher order. For qEDM, the $\pi N$ and $N\!N$ interactions are suppressed by $\alpha_{\mathrm{em}}$ and, as a consequence, we only need $\bar d_0$ and $\bar d_1$. Finally, for the $\chi$I sources all six interactions appear at LO. These differences between the various $\slashPT$ sources give rise to a different pattern of hadronic and nuclear $\slashPT$ moments.

\section{$P$- and $T$-violating moments} 
We first look at the electric dipole form factor (EDFF) of the nucleon. For $\tb$, $\chi$PT calculations have been performed at LO  \cite{Cre79,nEDMLO} and NLO \cite{nEDMNLO}. The isovector EDFF appears at LO due to a one-pion loop involving $\bar g_0$. The associated divergence is absorbed by the short-range counterterm $\bar d_1$. At NLO more loops appear but the results are not changed significantly \cite{nEDMNLO}. The isoscalar EDFF at LO is determined solely by the isoscalar short-range term $\bar d_0$, such that at this order the isoscalar EDFF is equal to the EDM. At NLO the first pion-loop contributions appear which induce the first energy-dependent correction to the isoscalar EDM, \textit{i.e.} the isoscalar nucleon Schiff moment. 

The nucleon EDFF for the dimension-six sources has been calculated within $\chi$PT in Refs \cite{nEDMNLO,Vri11a}. For the qCEDM the LO calculation is identical to that of the $\tb$ term, because loops involving $\bar g_1$ vanish at this order. For the qEDM and $\chi$I sources, the nucleon EDM is dominated solely by the short-range contributions $\bar d_0$ and $\bar d_1$. The first contributions to the Schiff moments appear at NNLO in the form of pion loops ($\chi$I sources) and short-range contributions (qEDM and $\chi$I sources). 

We conclude that for all sources the neutron $(d_n)$ and proton $(d_n)$ EDM depend on the two short-range interactions $\bar d_0$ and $\bar d_1$. For $\tb$ and qCEDM there is additional dependence on $\bar g_{0,1}$, but this dependence cannot be separated from the short-range contributions in a model-independent way. For all sources we expect $d_n$ and $d_p$ to be of the same order of magnitude such that measuring both of them gives no handle on separating the various $\slashPT$ sources. We can use the neutron EDM limit \cite{dnbound} to set limits on the different sources \cite{Vri11a}
\begin{eqnarray}\label{nEDMbounds}
\tb\simle 10^{-10},\qquad
\frac{ {\tilde \delta}}{M_{\slashTsub}^2},\,\frac{\delta}{M_{\slashTsub}^2} 
\simle \left(10^{5} \; {\rm GeV}\right)^{-2},\qquad
\frac{ w}{M_{\slashTsub}^2}
\simle \left(10^{6} \; {\rm GeV}\right)^{-2}.\label{dim6bound}
\end{eqnarray}
We see that EDM experiments probe scales beyond that of the LHC.
The nucleon Schiff moments could tell us more about the underlying source. For $\tb$ and qCEDM we expect the isovector Schiff moment to be (in appropriate units) of similar size as the nucleon EDM, while, for the qEDM and $\chi$I sources, the Schiff moments appear at NNLO. Unfortunately, there is no experimental access to the nucleon Schiff moments and we have to look at other observables in order to determine the dominant $\slashPT$ source. 

Triggered by the proposals to measure the EDMs of light nuclei in magnetic storage rings \cite{storageringexpts}, we turn our attention to these systems. $\chi$PT allows the calculations of these EDMs within the same framework as that used for the nucleon calculations. The wave functions of the bound states are calculated from modern high-quality phenomenological potentials. The nuclear $\slashPT$ properties can be calculated by inserting $\slashPT$ currents in the obtained nuclear wave function, or by perturbing the wave function with the $\slashPT$ $N\!N$ potential and inserting a $PT$ current. 
The LO $\slashPT$ currents \cite{Vri12} and potential \cite{TVpotential} are for all sources calculated from the interactions in Eq. \eqref{6Lecs} supplemented with the standard $PT$ $\chi$PT Lagrangian.

\begin{figure}[t]
\centering
\includegraphics[scale = 0.6]{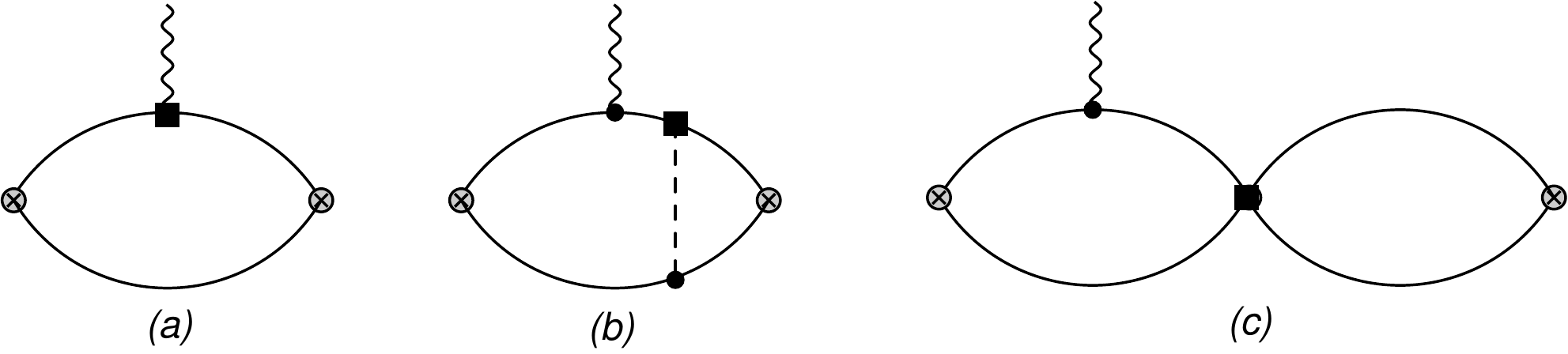}
\caption{Different diagrams contributing to the deuteron EDM. Solid, dashed, and wavy lines represent nucleons, pions, and photons, 
respectively. A square marks a $\slashPT$ interaction from Eq. \eqref{6Lecs},
other vertices representing LO $PT$ interactions. The crossed vertex represents the deuteron ground state. 
For simplicity only one possible ordering is shown here. The $PT$ $N\!N$ potential is inserted, but not explicitly shown, between the photon emission and $\slashPT$ interaction in diagrams (b) and (c)}
\label{deutLO}
\end{figure}

We begin our discussion of light nuclear EDMs with the deuteron. The LO diagrams contributing to the deuteron EDM ($d_d$) involving the interactions in Eq. \eqref{6Lecs} are shown in Fig. \ref{deutLO}. Although general light-nuclear EDMs depend on the six LECs in Eq. \eqref{6Lecs}, the deuteron is a special case due to its spin-isospin properties. The deuteron ground state is isoscalar and consists mainly of a ${}^3 S_1$ state. Any $\slashPT$ current needs to be isoscalar in order to contribute to $d_d$ so only the combination $d_n + d_p$ appears in diagram \ref{deutLO}(a). After perturbing the wave function in diagram \ref{deutLO}(b) by a $\slashPT$ one-pion exchange involving the vertex $\bar g_0$ or $\bar g_1$ the wave function will obtain, respectively, some ${}^1 P_1$ and ${}^3 P_1$ admixture. In the former case, because the $PT$ current coming from the proton charge is spin independent, it cannot bring the wave function from ${}^1 P_1$ to ${}^3 S_1$ and there is no contribution to $d_d$. Exactly the same argument explains why diagram \ref{deutLO}(c), involving the $\chi$I $\slashPT$ $N\!N$ interactions, does not give rise to $d_d$. In total, apart from higher-order corrections, $d_d$ is given by \cite{Vri12,Vri11b,Khrip,Liu04} 
\begin{equation}
d_{d} = d_{n}+d_{p}
-0.19\, \frac{\bar{g}_{1}}{\Fp} e\,\textrm{fm}. 
\label{dEDM}
\end{equation}
An important question is now which effect dominates this expression. Is it dominated by the one-body terms or by two-body dynamics involving $\bar g_1$? This depends on the source we are studying. For $\tb$, qEDM, and $\chi$I sources $\bar g_1$ is suppressed and $d_d$ is well approximated by $d_n + d_p$ \cite{Vri12,Vri11b}. For the qCEDM, however, $d_d$ is dominated by $\bar g_1$ and by NDA we estimate the nucleon EDMs to contribute at the $20\%$ level \cite{Vri12}. This implies that a measurement of $d_d$ significantly larger than $d_n + d_p$ would point towards physics beyond the SM in guise of a qCEDM. If experiments find $d_d \simeq d_n + d_p$, additional observables are needed to disentangle $\tb$ from the qEDM and $\chi$I sources.

In the latter case another $\slashPT$ property of the deuteron could play a role. Because the deuteron has spin $1$, apart from an EDM, it has a magnetic quadrupole moment (MQM). The deuteron MQM $(\mathcal M_d)$ can be calculated along completely similar lines as $d_d$ \cite{Liu:2012tra}, however, there are a few differences. First, diagram \ref{deutLO}(a) is not present because the nucleons themselves do not posses an MQM. Second, the coupling to the photon in diagrams \ref{deutLO}(b,c) is due to the nucleon magnetic moments instead of the charge. Because this interaction is spin dependent we find that, contrary to $d_d$, $\mathcal M_d$ does depend on $\bar g_0$ and $\bar C_{1,2}$. Third, for the qEDM, $\mathcal M_d$ is dominated by a $\slashPT$ $\pi N \g$ interaction not present in Eq. \eqref{6Lecs}. Details can be found in Refs. \cite{Vri11b,Liu:2012tra}, here we only comment on the interesting ratio $m_d \mathcal M_d/d_d$ where $m_d$ is the deuteron mass. For the $\tb$ term, we find this ratio to be large ($\Or(10)$) while for the dimension-six sources we find the ratio to be $\Or(1)$ or smaller. We conclude that $\mathcal M_d$ is sensitive to the $\tb$ term and that it would be interesting if it could be experimentally probed. 

Since so far there exists no method for accurately measuring $\mathcal M_d$, we turn to the EDMs of the helion $(d_{^{3}\mathrm{He}})$ and triton $(d_{^{3}\mathrm{H}})$. These systems do not have the spin-isospin filter the deuteron has, such that the EDMs depend on all six LECs in Eq. \eqref{6Lecs}. However, we find that the contributions from $\bar C_{1,2}$ are smaller than expected by power counting and they play no significant role. This might change for heavier systems where the binding energy per nucleon increases. Neglecting $\bar C_{1,2}$, $d_{^{3}\mathrm{He}}$ and $d_{^{3}\mathrm{He}}$ are found to be \cite{Vri12,Ste08}
\begin{eqnarray}
d_{^{3}\mathrm{He}}  &=& 
0.88\,d_n-0.05\,d_p 
-\left(0.15\,\frac{\bar{g}_{0}}{\Fp}+0.28\,\frac{\bar{g}_{1}}{\Fp}\right)
 e\,\mathrm{fm} \ ,
\label{hEDMqCEDM}
\\
d_{^{3}\mathrm{H}}  &=& 
-0.05\,d_n+0.90\,d_p
+\left(0.15\,\frac{\bar{g}_{0}}{\Fp}-0.28\,\frac{\bar{g}_{1}}{\Fp}\right)
 e\,\mathrm{fm}.
\label{tEDMqCEDM}
\end{eqnarray}
Again which of the interactions is important depends on the $\slashPT$ source. For $\tb$, the EDMs are dominated by the contribution from $\bar g_0$ and the nucleon EDMs appear at subleading order. The same holds for the qCEDM, but the tri-nucleon EDMs depend on $\bar g_1$ as well. For $\tb$ we expect the sum of $d_{^{3}\mathrm{He}}$ and $d_{^{3}\mathrm{He}}$ to lie close to the sum of the nucleon EDMs, $d_{^{3}\mathrm{He}}+ d_{^{3}\mathrm{H}} \simeq 0.94\,(d_n+d_p)$, while the difference is expected to deviate from $d_n-d_p$. For qCEDM we expect $d_{^{3}\mathrm{He}}+ d_{^{3}\mathrm{H}} \simeq 3 d_d >\!> d_n + d_p$. For qEDM and $\chi$I sources we expect $d_{^{3}\mathrm{He}}$  and $d_{^{3}\mathrm{He}}$ to lie close to, respectively, $d_n$ and $d_p$. We conclude that tri-nucleon EDM measurements could separate $\tb$ from the qEDM and $\chi$I sources.

\section{Conclusions}
We have calculated $\slashPT$ moments of the nucleon and light nuclei within the consistent framework of chiral perturbation theory. At the current experimental accuracies hadronic $\slashPT$ moments can be caused by the QCD $\tb$ term or by physics beyond the Standard Model which, at low energies, takes the shape of various $\slashPT$ operators of dimension six. We have shown that measurements of the nucleon, deuteron, helion, and triton EDMs (or a subset) could tell us a lot about the fundamental $\slashPT$ mechanism. In particular a large deuteron EDM with respect to the nucleon EDM would be a clear sign for physics beyond the SM. Additional information could be obtained by measurements of the nucleon Schiff moments or the deuteron magnetic quadrupole moment. 

\begin{acknowledgements}
I would like to thank Emanuele Mereghetti, Rob Timmermans, Bira van Kolck, C.-P. Liu, Ionel Stetcu,  Renato Higa, Claudio Maekawa, and Will Hockings for the enjoyable collaborations on various projects presented here. This research was supported by the
Dutch Stichting FOM under programs 104 and 114.

\end{acknowledgements}


\begin{thebibliography}{99}

\bibitem{'tHooft:1976up}
  G. 't Hooft,
 Phys. Rev. Lett. {\bf 37}, 8 (1976).

\bibitem{dnbound}
C. A. Baker {\it et al.}, 
Phys. Rev. Lett. {\bf 97}, 131801 (2006). 

\bibitem{Cre79}
R.J. Crewther, P. Di Vecchia, G. Veneziano, and E. Witten,
Phys. Lett. B {\bf 88}, 123 (1979); {\bf 91}, 487(E) (1980). 

\bibitem{storageringexpts}
F. J. M. Farley {\it et al.},
Phys. Rev. Lett. {\bf 93}, 052001 (2004);
C. J. G. Onderwater, 
J. Phys. Conf. Ser. {\bf 295}, 012008 (2011);
Y.~K.~Semertzidis,
  J.\ Phys.\ Conf.\ Ser.\  {\bf 335}, 012012 (2011). 

\bibitem{dim6origin}
W.~Buchm\"uller and D.~Wyler,
Nucl. Phys. {\bf B268}, 621 (1986);
B.~Grzadkowski, M.~Iskrzynski, M.~Misiak, and J.~Rosiek,
JHEP {\bf 1010}, 085 (2010).


\bibitem{Fujikawa:1979ay}
  K.~Fujikawa,
  Phys.\ Rev.\ Lett.\  {\bf 42}, 1195 (1979); V. Baluni, Phys. Rev. D {\bf 19}, 2227 (1979).
  

\bibitem{Rujula}
A.~De R\'ujula, M. B.~Gavela, O.~P\`ene, and F. J.~Vegas,
Nucl. Phys. {\bf B357}, 311 (1991).

\bibitem{Weinberg:1989dx}
S.~Weinberg,
Phys.\ Rev.\ Lett.\  {\bf 63}, 2333 (1989).

  
\bibitem{Pospelov:2005pr}
M. Pospelov and A. Ritz,
Ann. Phys.\ {\bf 318}, 119 (2005).

\bibitem{NDA}
A. V. Manohar and H. Georgi,
Nucl. Phys. B {\bf 234}, 189 (1984).  

\bibitem{BiraEmanuele} 
E. Mereghetti, W. H. Hockings, and U. van Kolck,
Ann. Phys. {\bf 325}, 2363 (2010).

\bibitem{TVpotential}
C. M. Maekawa, E. Mereghetti, J. de Vries, and U. van Kolck,
Nucl. Phys. A {\bf 872}, 117 (2011). 

\bibitem{dim6} 
J. de Vries, E. Mereghetti, R. G. E. Timmermans, and U. van Kolck,
in preparation.

\bibitem{Vri12}
J.~de Vries, R.~Higa, C.~-P.~Liu, E.~Mereghetti, I.~Stetcu, 
R.~G.~E.~Timmermans, and U.~van Kolck,
Phys.\ Rev.\ C {\bf 84}, 065501 (2011).

\bibitem{Jenkins:1990jv}
E. E. Jenkins and A. V. Manohar,
Phys.\ Lett.\  B {\bf 255}, 558 (1991).

\bibitem{nEDMLO}
A. Pich and E. de Rafael, 
Nucl. Phys. B {\bf 367}, 313 (1991);
S. D. Thomas,
Phys.\ Rev.\  D {\bf 51}, 3955 (1995);
W. H. Hockings and U. van Kolck, 
Phys. Lett. B {\bf 605}, 273 (2005).

\bibitem{nEDMNLO}
S.~Narison,  Phys.\ Lett.\ B {\bf 666}, 455 (2008);
K. Ottnad, B. Kubis, U.-G. Mei{\ss}ner, and F.-K. Guo,
Phys. Lett. B {\bf 687}, 42 (2010);
E. Mereghetti, J. de Vries, W. H. Hockings, C. M. Maekawa, and U. van Kolck,
Phys. Lett. B {\bf 696}, 97 (2011).

\bibitem{Vri11a}
J. de Vries, E. Mereghetti, R. G. E. Timmermans, and U. van Kolck,
Phys. Lett. B {\bf 695}, 268 (2011).
  

\bibitem{Vri11b}
J. de Vries, E. Mereghetti, R. G. E. Timmermans, and U. van Kolck,
Phys. Rev. Lett. {\bf 107}, 091804 (2011).

\bibitem{Khrip}
I. B. Khriplovich and R. V. Korkin, 
Nucl. Phys. {\bf A665}, 365 (2000).

\bibitem{Liu04}
C.-P. Liu and R. G. E. Timmermans, 
Phys. Rev. C {\bf 70}, 055501 (2004).

\bibitem{Liu:2012tra}
  C.~-P.~Liu, J.~de Vries, E.~Mereghetti, R.~G.~E.~Timmermans and U.~van Kolck,
  Phys.\ Lett.\ B {\bf 713}, 447 (2012).
  
\bibitem{Ste08}
I. Stetcu, C.-P. Liu, J. L. Friar, A. C. Hayes, and P. Navr\'{a}til, 
Phys. Lett. B {\bf 665}, 168 (2008).  

\end{thebibliography}
\end{document}